\documentclass[runningheads]{llncs}

\usepackage{epsfig,endnotes,url}

\usepackage{graphicx}
\usepackage{float}
\usepackage{multirow}
\usepackage{algorithm}
\usepackage{algorithmicx}
\usepackage{algpseudocode}
\usepackage{xspace}

\usepackage{xparse}
\usepackage{balance}
\usepackage{tikz}
 
\usepackage{paralist}

\usepackage{amsmath} 
\usepackage{xcolor}  
\usepackage{comment} 
\usepackage{wrapfig,lipsum,booktabs} 

\newcommand{\etal}{\textnormal{et al.}\xspace}

\newcommand{\eg}{\textnormal{e.g.,}\xspace}

\newcommand{\subhead}[1]{\vspace{1pt}\noindent\textbf{#1.}}

\newenvironment{compactlist}
  {\begin{itemize}
  \setlength{\itemsep}{0pt}
  \setlength{\parskip}{0pt}}
  {\end{itemize}}

\newenvironment{compactlistn}
  {\begin{enumerate}
  \setlength{\itemsep}{0pt}
  \setlength{\parskip}{0pt}}
  {\end{enumerate}}

\newcommand{\tuple}[1]{\left \langle #1 \right \rangle}

\newcommand{\bfhead}[1]{\vspace {3pt}\noindent{\textbf{#1}}}

\usepackage{booktabs} 
\usepackage{pdfpages}

\begin{document}
\title{One-Time Programs made Practical}

\author{Lianying Zhao\inst{1} \and
Joseph I. Choi\inst{2} \and
Didem Demirag\inst{3} \and
Kevin R. B. Butler\inst{2} \and
Mohammad Mannan\inst{3} \and
Erman Ayday\inst{4} \and
Jeremy Clark\inst{3}}
\authorrunning{L. Zhao et al.}

\institute{University of Toronto, Toronto ON, Canada \and
University of Florida, Gainesville FL, USA \and
Concordia University, Montreal QC, Canada \and
Case Western Reserve University, Cleveland OH, USA
%
}

\maketitle

\begin{abstract}
A one-time program (OTP) works as follows: Alice provides Bob with the implementation of some function. Bob can have the function evaluated exclusively on a single input of his choosing. Once executed, the program will fail to evaluate on any other input. State-of-the-art one-time programs have remained theoretical, requiring custom hardware that is cost-ineffective/unavailable, or confined to ad-hoc/unrealistic assumptions. To bridge this gap, we explore how the Trusted Execution Environment (TEE) of modern CPUs can realize the OTP functionality. 
Specifically, we build two flavours of such a system: in the first, the TEE directly enforces the one-timeness of the program; in the second, the program is represented with a garbled circuit and the TEE ensures Bob's input can only be wired into the circuit once, equivalent to a smaller cryptographic primitive called one-time memory. These have different performance profiles: the first is best when Alice's input is small and Bob's is large, and the second for the converse. 
\end{abstract}



\section{Introduction}

Consider the well-studied scenario of secure two-party computation:
Alice and Bob want to compute a function on their inputs, but they do not want
to disclose these inputs to each other (beyond what can be inferred from the
output of the computation). This is traditionally handled by an interactive
protocol between Alice and Bob.\footnote{Hazay and Lindell~\cite{HL10}
give a thorough treatment of interactive two-party protocols.}
In this paper, we instead study a non-interactive protocol as follows:
Alice prepares a device for Bob with the function and her input included;
once Bob receives this device from Alice, he supplies his input and learns
the outcome of the computation. The device will not reveal the outcome for
any additional inputs (thus, a one-time program~\cite{goldwasserOTP}).
Alice might be a company selling the device in a retail store, and Bob the
customer; the two never interact directly. By using the device offline, Bob is
assured that his input remains private.

To build a one-time program (OTP), we use the Trusted Execution Environment (TEE),
a hardware-assisted secure mode on modern processors, where execution integrity and
secrecy are ensured~\cite{mccune2009}, with qualities that include platform state binding
and protection of succinct secrets.\footnote{Further explanation is provided in Appendix~\ref{sec:tee}.}
TEEs may appear to offer a trivial solution to OTPs; however, complexities arise due to
Bob's physical possession of the device and, more importantly, performance issues.
We propose two configurations for one-time programs built on TEEs: (1) deployed directly in the TEE,
and (2) deployed indirectly via TEE-backed one-time memory (OTM)~\cite{goldwasserOTP}
and garbled circuits~\cite{yao_gc}
outside of the TEE. OTMs hold two keys, only one of which gets revealed
(dependent on its input); the other is effectively destroyed.

\paragraph{Contributions.} Our system, built using Intel Trusted Execution Technology (TXT)~\cite{TXT} and Trusted Platform Module (TPM)~\cite{tpmspec} as the TEE, is available today (as opposed to custom OTP/OTM implementations using FPGA~\cite{fpga-otp}, PUF~\cite{puf}, quantum mechanisms~\cite{q-otp}, or online services~\cite{moneyOTP}) and could be built for less than \$500.\footnote{As an example, Intel STK2mv64CC, a Compute Stick that supports both TXT and TPM, was priced at \$499.95 USD on Amazon.com (as of September 2018).} 

We propose and implement the following OTP variants, considering that TPM-sealing\footnote{A state-bound cryptographic operation performed by the TPM chip, like encryption.} or encrypting data is time-consuming. 
\vspace{-5pt}
\begin{compactlist}
\item \emph{TXT-only} seals/unseals Alice's input directly, and performance is thus sensitive to Alice's input size. Bob's input is entered in plaintext and processed in TXT after he has received the device.  
\item \emph{GC-based} converts the logic into garbled circuit, where number of key pairs is determined by Bob's input size. Key pairs are encrypted/decrypted with a master key (MK). 
This way, the performance is largely determined by Bob's input size. Upon receiving the device, he does the one-time selection of key pairs in TXT to reflect his input. Thereafter, evaluation of the garbled circuit can be done on any machine with the selected keys.
\end{compactlist}
\vspace{-5pt}
To illustrate the generality of our solution, we also map the following application into our proposed OTP paradigm: a company selling devices that will perform a private genomic test on the customer's sequenced genome. For this use case, in one of our two variants (TXT-only), a company can initialize the device in 5.6 seconds and a customer can perform a test in 34 seconds.


\vspace{-10pt}
\section{Preliminaries}
\label{sec:prelim}

\vspace{-5pt}
\subsection{One-Time Program Background}
\label{sec:otpbg}
\vspace{-5pt}
A one-time program can be conceived of as a non-interactive version of a two party computation: $y=f(a,b)$ where $a$ is Alice's private input, $b$ is Bob's, $f$ is a public function (or program), and $y$ is the output. Alice hands to Bob an implementation of $f_a(\cdot)$ which Bob can evaluate on any input of his choosing: $y_b=f_a(b)$. Once he executes on $b$, he cannot compute $f_a(\cdot)$ again on a different input. For our practical use-case, we conceive of OTPs with less generality as originally proposed by Goldwasser \etal~\cite{goldwasserOTP}; essentially we treat them as one-time, non-interactive programs that hide Alice and Bob's private inputs from each other without any strong guarantees on $f$ itself. Note with a general compiler for $f$ (which we have for both flavours of our system), it is easy but inefficient to keep $f$ private.\footnote{Essentially, one would define a very general function we might call $\mathsf{Apply}$ that will execute the first input variable on the second: $y=\mathsf{Apply}(f,b)=f(b)$. Since $f$ is now Alice's private input, it is hidden. The implementation of $\mathsf{Apply}$ might be a universal circuit where $f$ defines the gates' logic --- in this case $\mathsf{Apply}$ would leak (an upper-bound on) the circuit size of $f$ but otherwise keep $f$ private.}
Additional background on one-time programs is provided in Appendix~\ref{appendix:otp}.

\subsection{Threat Model and Requirements}
\label{sec:threat}

We informally consider an OTP to be secure if the following properties are achieved: (1) Alice's input $a$ is confidential from Bob; (2) Bob's input $b$ is confidential from Alice, and (3) no more than one $b$ can be executed in $f(a,b)$ per device.
We argue the security of our two systems in Section~\ref{sec:security} but provide a synopsis here first. Property 3 is enforced through a trusted execution environment, either directly (TXT-only variant in Section~\ref{sec:txtonly}) or indirectly via a one-time memory device (GC-based TXT in Section~\ref{sec:gc}) as per the Goldwasser \etal construction. Given Property 3, we consider Property 1 to be satisfied if an adversary learns at most negligible information about $a$ when they choose $b$ and observe $\tuple{\mathsf{OTP},f(a,b),b}$ as opposed to simply $\tuple{f(a,b),b}$, where $\mathsf{OTP}$ is the entire instantiation of the system, including the TPM-sealed memory and system details (and for the GC-variant: the garbled circuit and keys revealed through specifying $b$). Property 2 is achieved by being provisioned an offline device that can compute $f_a(b)$ without any interaction with Alice. There is a possibility that the device surreptitiously stores Bob's input and tries to leak it back to Alice. We discuss this systems-level attack in Section~\ref{sec:security}. In Appendix~\ref{app:adaptive}, we address a subtle adaptive security attack applicable to the original OTP system.  

The seleciton of TEE has to reflect the aforementioned Properties 1 and 3. Property 3 is achieved by stateful (recording the one-time state) and integrity-protected (enforcing one-timeness) execution, which is the fundamental purpose of all today's TEEs.
Moreover, both Properties 1 and 3 mandate no information leakage, which can occur through either software or physical side-channels. We choose Intel TXT, primarily because of its
\emph{exclusiveness}, which means: TXT occupies the entire system when secure execution is started and no other code can run in parallel. This naturally avoids all software side-channels, an advantage over non-exclusive TEEs. We do consider using non-exclusive TEEs as future exploration when the challange of software side-channels has been overcome, e.g., for Intel SGX, the (recent) continually identified side-channel attacks, such as Foreshadow~\cite{foreshadow}, branch shadowing~\cite{branchshadow}, cache attacks~\cite{sgxcache}, and more; for ARM TrustZone, there have been TruSpy~\cite{truspy}, Cachegrab~\cite{cachegrab}, etc.
They all point to the situation when trusted and untrusted code run on shared hardware.

The known physical side-channels can also be mitigated in the setting of our OTP, i.e., DMA attacks are impossible if I/O protection is enable (by the chipset), and the cold-boot attack~\cite{cold-boot} can be avoided if we choose computers with RAM soldered on the motherboard (cannot be removed to be mounted on anohther machine, see Section~\ref{sec:security}). For a detailed comparison with other TEEs, refer to Appendix~\ref{sec:tee}).

We strive for a reasonable, real-world threat model where we mitigate attacks introduced by our system but do not necessarily resolve attacks that apply broadly to practical security systems. Specifically, we assume:
\begin{compactlist}
\item Alice is monetarily driven or at least curious to learn Bob's input, while Bob is similarly curious to learn the algorithm of the circuit and/or re-evaluate it on multiple inputs of his choice.
\item We assume Alice produces a device that can be reasonably assured to execute as promised (disclosed source, attestation quotes over an integral channel, and no network capabilities). 
\item We assume that Alice's circuit (including the function and her input) actually constitutes the promised functionality (\eg is a legitimate genomic test). 
\item We assume the sound delivery of the device to Bob. We do not consider devices potentially subverted in transit which applies to all electronics~\cite{nsa}. 
\item Both Alice and Bob have to trust the hardware manufacturer (in our case, Intel and the TPM vendor) for their own purposes. 
Alice trusts that the circuit can only be evaluated once on a given input from Bob, while 
Bob trusts that the received circuit is genuine and the output results are trustworthy.
\item Bob has only bounded computational power, 
and may go to some lab effort, such as tapping pins on the motherboard and cloning a hard drive, but not efforts as complicated as imaging a chip~\cite{tpmimg,cpuimg1,cpuimg2}.
\item Components on the motherboard cannot be manipulated easily (\eg forwarding TPM traffic from a forged chip to a genuine one by desoldering). 
\end{compactlist}

\vspace{-5pt}
\subsection{Intel TXT and TPM}
\label{sec:tpm}
Intel Trusted Execution Technology (TXT) is also known as ``late launch'', for its capability to launch secure execution at any point, occupying the entire system.
When the CPU enters the special mode of TXT, all current machine state is discarded/suspended and a fresh secure session is started, hence its exlusiveness, as opposed to sharing hardware with untrusted code.

\subhead{Components}
TXT relies on three mandatory hardware components to function: 
\begin{inparaenum}[a)]
\item CPU. The instruction set is extended with a few new instructions for the management of TXT execution.
\item Chipset. The chipset (on the motherboard) is respondible for enforcing I/O protection such that the specified range of I/O space is only accesible by the protected code in TXT; and
\item TPM. Trusted Platform Module~\cite{tpmspec} is a microchip, serving as the secure storage (termed Secure Element).  Its \emph{PCR} (Platform Configuration Register) is volatile storage containing the machine state, in the form of concatenated hash values. There are also multiple PCRs for different purposes. On the TPM, there is also non-volatile storage (termed \emph{NVRAM}), allocated in the unit of \emph{index} of various sizes. Multiple indices can be defined depending on the capacity of a specific TPM model.
\end{inparaenum}

\subhead{Measured launch}
A provisioning stage is always involved where the platform is assumed trusted and uncompromised.  A piece of code is measured (similar to hashing) and the measurements are stored in certain TPM NVRAM indices as policies.
Thereafter (in our case in the normal execution mode with Bob), the program being loaded is measured and compared with the policies stored in TPM. The system may then abort execution if mismatch is detected, or otherwise proceed.
This process is enforced by the CPU. 

\subhead{Machine state binding}
As run-time secrecy (secret in use) is ensured by measured launch and I/O isolation, we also need secrecy for stored data (secret at rest). Alice's input should not be learned by Bob when the device is shipped to him.
From the start of TXT execution, each stage measures the next stage's code and \emph{extends} the hash values as measurement to the PCR (concatenated and hashed with the existing value). This way, the measurements are chained, and at a specific time the PCR value reflects what has been loaded before. The root of this chained trust is the measured launch.

Such chained measurements (in PCRs) can be used to derive the key for data encryption, so that only when a desired software stack is running can the protected data be decrypted. This cryptographic operation performed by the TPM is termed \emph{sealing}. A piece of data sealed under certain PCRs can only be unsealed under the same PCRs, hence bound to a specific machine state. The sealed data (ciphertext) can be stored anywhere depending on its size. It is noteworthy to mention that there exists a distinct equivalent of sealing which, instead of just encryption, stores data in a TPM NVRAM index and binds its access to a set of PCRs. As a resuilt, without the correct machine state, the NVRAM index is completely inaccessible (read/write) and thus replaying the ciphertext is prevented. We term it \emph{PCR-bound NVRAM sealing} in this paper and use it for our OTP prototype implementation.

\vspace{-5pt}
\section{Related Work}
\label{sec:relwork}
\vspace{-5pt}

In the original one-time program paper by Goldwasser \etal~\cite{goldwasserOTP}, OTM is left as a theoretical device.  In the ensuing years, there have been some design suggestions based on quantum mechanisms~\cite{q-otp}, physically unclonable functions~\cite{puf}, and FPGA circuits~\cite{fpga-otp}. 
\begin{inparaenum}[(\bgroup\bfseries a\egroup)]
\item J\"{a}rvinen \etal~\cite{fpga-otp} provide an FPGA-based implementation for GC/OTP, with a GC evaluation of AES, as an example of a complex OTP application.  They conclude that although GC/OTP can be realized, their solution should be used only for ``truly security-critical applications'' due to high deployment and operational costs. They also provide a cryptographic mechanism for protecting against a certain adaptive attack with one-time programs (see Appendix~\ref{app:adaptive}); it is tailored for situations where the function’s output size is larger than the length of a special holdoff string stored at each OTM. 
\item Kitamura \etal~\cite{moneyOTP} realize OTP without OTM by proposing a distributed protocol, based on secret sharing, between non-colluding entities to realize the `select one key; delete the other key' functionality. This introduces further interaction and entities. Our approach is in the opposite direction: removing all interaction (other than transfer of the device) from the protocol. 
\item Prior to OTP being proposed, Gunupudi and Tate~\cite{tpm-otm} proposed count-limited private key usage for realizing non-interactive oblivious transfer using a TPM. Their solution requires changes in the TPM design (due to lack of a TEE). In contrast, we utilize unmodified TPM 1.2.
\item In a more generalized setting, ICE~\cite{ice} and Ariadne~\cite{ariadne} consider the state continuity of any stateful program (including N-timeness) in the face of unexpected interruption, and propose mechanisms to ensure both rollback protection and usability (i.e., liveness).  We solve the specific problem of one-timeness/N-timeness, focusing more on how to deal with input/output and its implication on performance. We do sacrifice liveness (i.e., we flip the one-timeness flag upon entry and thus the program might run zero time if crashed halfway). We believe their approaches can be applied in conjunction with ours.
\end{inparaenum}



\vspace{-10pt}
\section{System 1: TXT-only}
\label{sec:txtonly}
\vspace{-5pt}

\paragraph{Overview.} 
In the first system, we propose to achieve one-timeness by running the protected program in TEE only once (relying on logic integrity) and storing its persistent state (e.g., the one-time indicator) in a way that it is only accessible from within the TEE. 
To eliminate information leakage from software side-channels, we have chosen Intel TXT for its exclusiveness (i.e., no other software in parallel).\footnote{We consider various TEE options and provide our reasons for choosing Intel TXT to instantiate the TEE in Appendix~\ref{sec:tee}.} We hence name this design \emph{TXT-only}.

To achieve minimal TCB (Trusted Computing Base) and simplicity, we choose native C programming in TXT (as opposed to running an OS/VM). Therefore, for one-time programs that have existing implementation in other languages, per-application adaptation is required (cf.\ similar porting effort is needed for the GC-based variant in Section~\ref{sec:gc}). For further consideration of the porting effort required, refer to Appendix~\ref{appendix:port}. For new programs, this may not introduce extra effort.
\vspace{-10pt}
\paragraph{Design.} We briefly describe the components and workflow of the TXT-only system as follows.
A one-time indicator (flag) is sealed into the PCR-bound TPM NVRAM to prevent replay attacks. The indicator is checked and then flipped upon entry of the OTP. Without network connection, the device shipped to the client can no longer leak any of the client's secrets to the vendor. Therefore, only the vendor's secret input has to be protected. We TPM-seal the vendor input on hard drive for better scalability, and there is no need to address replay attacks for vendor input as one-timeness is already enforced with the flag.

The OTP program is loaded by the Intel official project \emph{tboot}~\cite{tboot} and GRUB. It complies with the Multiboot specification~\cite{multiboot}, and for accessing TPM, we reuse part of the code from tboot, and develop our own functions for commands that are unavailable elsewhere, e.g., reading/writing indices with PCR-bound NVRAM sealing. Since we do not load a whole OS into TXT with tboot, we cannot use OS services for disk I/O access; instead, we implement raw PATA (Parallel ATA, a legacy interface to the hard drive, compatible mode with SATA) logic and directly access disk sectors with DMA (Direct Memory Access). In the \emph{provisioning mode}, the OTP program performs a one-time setup, such as initiating the flag in NVRAM, sealing (overwriting) Alice's secret, etc. Once the \emph{normal execution mode} is entered, the program will refuse to run a second time.

\subhead{Memory exposure}
As an optional feature for certain computers with swappable RAM, we expose the unsealed vendor input in very small chunks during execution. For example, if the vendor input has 100 records, we would unseal one record into RAM each iteration for processing the whole user input. This way, in case of the destructive cold boot attack, the adversary only learns one-hundredth of the vendor's secret, and no more attempts are possible (the indicator is already updated). 

\vspace{-10pt}
\subsection{TXT-only provisioning/evaluation}
\begin{figure}[t]
\centering
\includegraphics[width=0.6\textwidth]{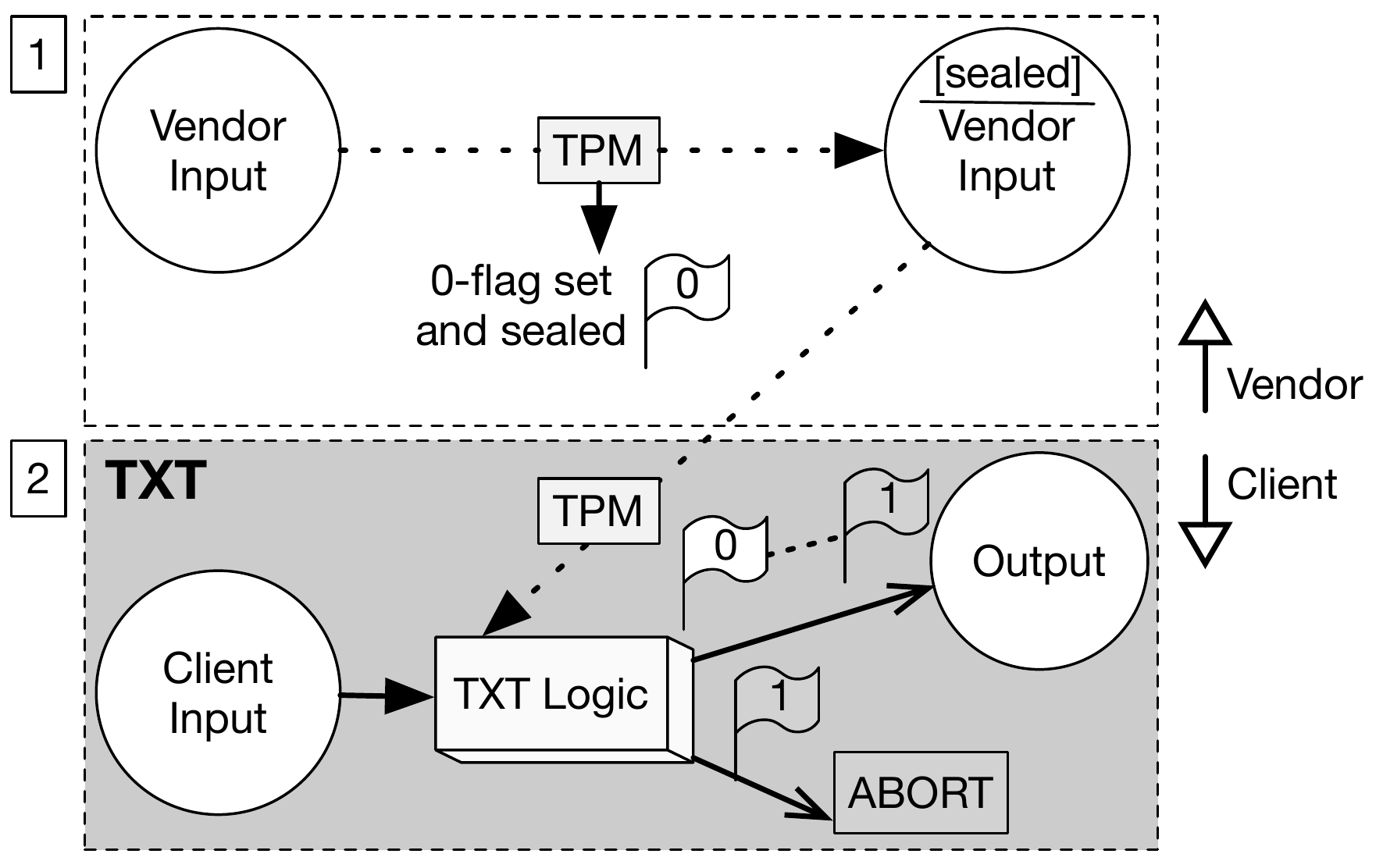}
\vspace{-5pt}
\caption{
Our realization of OTPs spans two
phases when relying on TXT alone for the entire computation.
Alice is active only during phase 1; Bob only during phase 2.}
\label{fig:txtonlyphases}
\vspace{-15pt}
\end{figure}

\vspace{-5pt}
Figure~\ref{fig:txtonlyphases} gives an overview of~\emph{TXT-only}, illustrating the initial provisioning
by Alice and evaluation of the function upon delivery to Bob. Note that what is delivered to Bob is the entire computer in our prototype (laptop or barebone like Intel NUC). \looseness=-1

\subhead{Provisioning at Alice's site}
At first, Alice is tasked with setting up the box, which will be
delivered to Bob. Alice performs the following:
(1) Write the integrity-protected payload/logic in C adapted to the
native TXT environment, e.g., static-linking any external libraries
and reading input data in small chunks. We may refer to it as the TXT program thereinafter.
(2) In the provisioning mode, initialize the flag to 0 and seal.\footnote{A flag is more straightforward to implement than a TPM monotonic counter, thanks to the PCR-bound NVRAM sealing, whereas a counter
would involve extra steps (such as attesting to the counterAuth password).}
The one-timeness flag is stored with the PCR-bound NVRAM sealing. 
Instead of depending on a password and regular sealing, this is like
stronger access-controlled ciphertext.
(3) Seal Alice's input onto the hard drive. 

\subhead{Evaluation at Bob's site}
After receiving the computation box from Alice,
Bob performs the following:
(1) Place the file with Bob's input on the hard drive.
(2) Load the TXT program in normal execution mode, which will read in Bob's input
and unseal Alice's input to compute on.
(3) Receive the evaluation result (e.g., from the screen or hard drive).
As long as it is Bob's first attempt to run the TXT program,
the computation will be permitted and the result will be returned to Bob.
Otherwise, the TXT program will abort upon loading in step (2),
as shown in Figure~\ref{fig:txtonlyphases}.

\vspace{-10pt}
\section{System 2: GC-based}
\label{sec:gc}
\vspace{-5pt}
As seen in our TXT-only approach to OTP (System 1) the data processing for protection is only applied to Alice's
input (with either sealing/unsealing or encryption/decryption), and Bob's input is always exposed
in plaintext due to the machine's physical possession by Bob. Intuitively, we may think that
it is a good choice when Alice's input is relatively small regardless of Bob's input size. However, there might be other
applications where Alice's input is substantially larger and become the performance bottleneck.
Is there a construction that complements TXT-only and is less sensitive to Alice's input size? 
The answer may lie in garbled circuits. During garbled circuit execution, randomly generated strings (or keys) are 
used to iteratively unlock each gate until arriving at the final output. 
Alice's input (size) is only ``reflected'' in the garbled circuit (assumed not trivially invertable~\cite{goldwasserOTP}), and the key pairs (whose number is determined by Bob's input size, not to do with Alice's) are sealed/encrypted, hence insensitive to Alice's input size. Details of garbled circuits and their use are explained in Appendix~\ref{appendix:gc}.

To adapt garbled circuits
for OTP, key generation and key selection steps are separated. As long as we limit
key selection to occur a single time, and the unchosen key of each key pair is
never revealed, we can prevent running a particular circuit on a different input.
To prevent keys from being selected more than once, we need to instantiate a one-time memory
(OTM), which reveals the key corresponding to each input bit and effectively destroys (or its equivalent)
the unchosen key in the key pair. OTM is left as a theoretical device in the
original OTP paper~\cite{goldwasserOTP}.
We realize it using Intel TXT and the TPM.
As in System 1, we seal a one-time flag into the PCR-bound TPM NVRAM,
and minimize the TXT logic to just handle key selection,
in preparation for GC execution.
Alice will seal (in advance) key pairs for garbling Bob's inputs.
Bob may then boot into TXT to receive the keys corresponding to his input.
When Bob reads a key off the device (say for input bit 0), the corresponding
key (for input bit 1) is erased.\footnote{Unselected keys remain sealed, if never unsealed it serves as cryptographic deletion.}
By instantiating an OTM in this manner, we can replace interactive oblivious transfer (OT)
and perform the rest of the garbled circuit execution offline, passing
key output from trusted selection.
By combining TXT and garbled circuits in this way,
sealing complexity is now tied to Bob's inputs.
We name this alternate construction~\emph{GC-based} (System 2).

\subhead{Performance overhead with TPM sealing}
According to our measurement, each TPM sealing/unsealing operation takes about 500ms, and therefore 1 GB of key pairs would need about 1000 hours, which is infeasible. Instead, we generate a random number as an encryption key (MK) at provisioning time and
the GC key pairs are encrypted with MK. We only seal MK. This way, MK becomes
per-deployment, and reprovisioning the system will not make the sealed key pairs
reusable due to the change of MK (i.e., the old MK is replaced by the new key).
Note that we could also apply the same approach to TXT-only (i.e., encrypting
Alice's input with MK and sealing only MK), if needed by the application.

\subhead{Memory exposure}
Similarly to the TXT-only OTP, our GC-based approach can also optionally
adapt to address the cold-boot attack. MK becomes a single point of failure if exposed in such memory
attacks, i.e., all key pairs can be decrypted and one-timeness is lost.
As with TXT-only, for smaller-sized client input, we can seal the key pairs directly and only unseal into RAM
in small chunks. \looseness=-1 

\begin{figure}[t]
\centering
\includegraphics[width=0.6\textwidth]{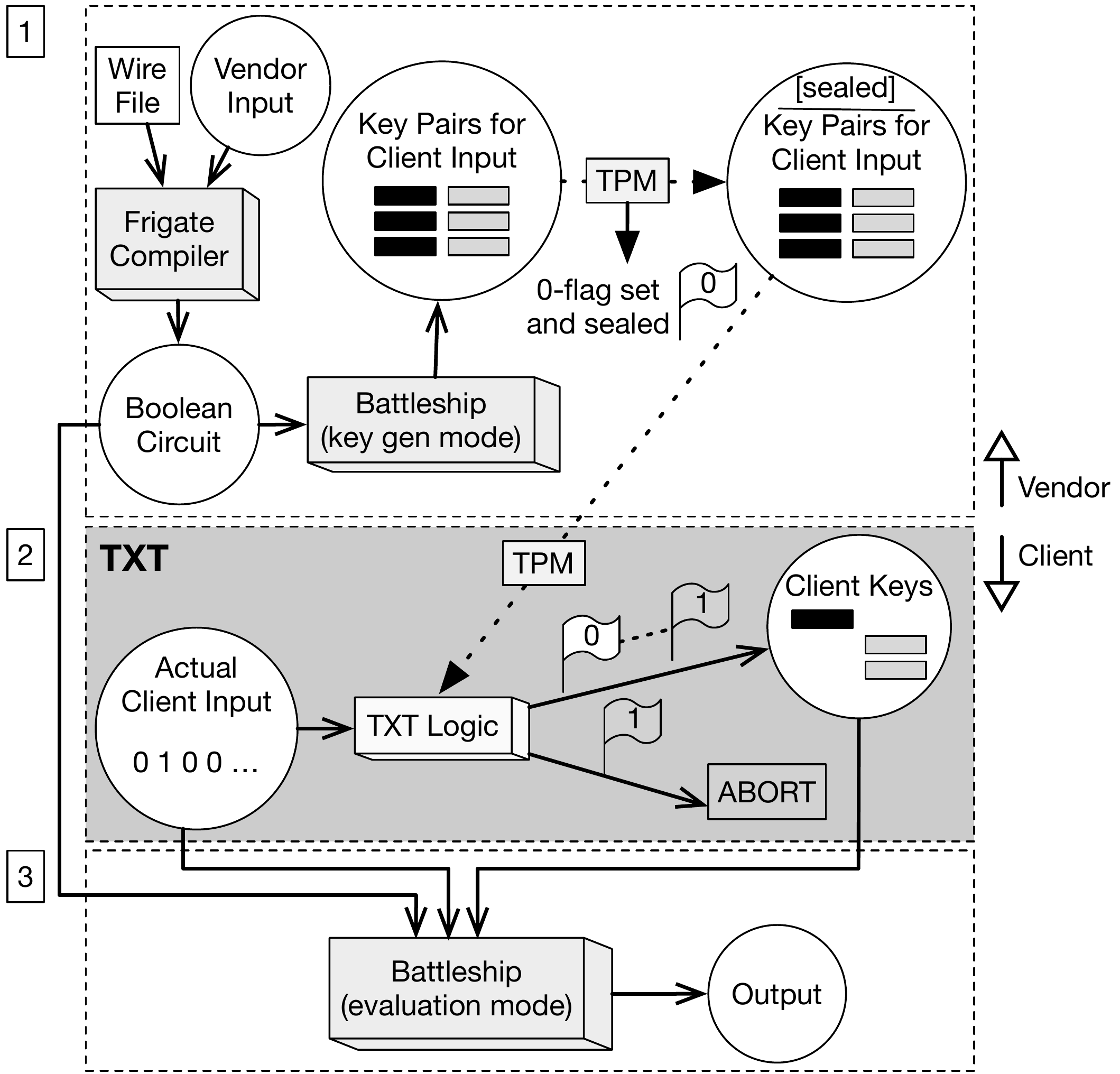}
\caption{
In our GC-based approach to OTP, Alice generates key pairs
and seals them. Bob unseals the keys that correspond
to his input and locally evaluates the function. For details about Frigate and Battleship, see Appendix~\ref{sec:frigmods}}
\label{fig:phases}
\vspace{-10pt}
\end{figure}

\vspace{-5pt}
\subsection{Implementation}
We use the Boolean circuit compiler~\emph{Frigate}~\cite{frigate} to implement the garbled circuit components of~\emph{GC-based}.
The interpreter and execution functionalities of~\emph{Frigate} are separately referred to as~\emph{Battleship}.
For our purposes, we split~\emph{Battleship} execution into two standalone phases: a key pair generation phase
({\tt gen}) and a function evaluation phase ({\tt evl}).
The motivation behind our choice of~\emph{Frigate} and our specific 
modifications to~\emph{Battleship} are detailed in Appendix~\ref{sec:frigmods}.

Our GC-based approach to OTP relies on TXT for trusted key selection and  leaves the computation
for garbled circuits, as shown in Figure~\ref{fig:phases}.  In our setting, Alice represents the vendor and Bob represents the client.

\subhead{Provisioning at Alice's site}
Alice sets up the OTP box by doing the following:
(1) Initialize flag to 0 and seal in the TXT program's provisioning mode.
(2) Write and compile, using {\em Frigate}, the wire program (.wir),
together with Alice's input, into the circuit.\footnote{
The wire program may be written and compiled on a separate
machine from that which will be shipped to Bob. If Alice chooses
to use the same machine, the (no longer needed) raw wire code and
{\em Frigate} executable should be removed from the box before 
provisioning continues.}
(3) Load the compiled .mfrig and .ffrig files, vendor's input, and the {\em Battleship}
executable onto the box.
(4) Write the TXT program (for key selection) in the same way as in TXT-only.
(5) Run {\em Battleship} in key-generation mode to generate
the $k^0_i$ and $k^1_i$ key-pairs corresponding to each of the $i$ bits of
Bob's input. These are saved to file.
(6) Seal the newly generated key pairs onto
the hard-drive in provisioning mode of the TXT program.
Alice is able to generate the correct number of key pairs, since garbled
circuit programs take inputs of a predetermined size, meaning Alice knows
the size of Bob's input.
Costly sealing of all key pairs could be switched out for sealing of the master key (MK) used to encrypt
the key pairs.

\subhead{Evaluation at Bob's site}
Bob, upon receiving the OTP box from Alice, performs the following steps to evaluate the function on his input:
(1) Place the file with Bob's input bits on the hard drive.
(2) Load the TXT program in normal (non-provisioning) mode for key selection.
(3) Receive selected keys corresponding to Bob's input bits; these are output to disk in plaintext.
As long as it is Bob's first attempt to select keys,
the TXT program will return the keys corresponding to Bob's input. 
Otherwise,
the TXT program will abort upon loading in step (2),
as shown in Figure~\ref{fig:phases}.
After Bob's inputs have been successfully garbled
(or converted into keys) and saved on the disk,
Bob can continue with the evaluation properly.
TXT is no longer required.
(4) Reboot the system into the OS (e.g., Ubuntu).
(5) Launch {\em Battleship} in circuit-evaluation mode.
(6) Receive the evaluation result from {\em Battleship}.
When {\em Battleship} is launched in circuit-evaluation mode,
the saved keys corresponding to Bob's input are read in.
{\em Battleship} also takes
vendor input (if not compiled into the circuit) before processing the garbled circuit.
The Boolean circuit is read in from the .mfrig and .ffrig files produced by
{\em Frigate}.
Evaluation is non-interactive and offline.
The evaluation result is available only to Bob.



\vspace{-10pt}
\section{Case Study}
\label{sec:case}
\vspace{-5pt}
We apply our proposed systems on a concrete use case based on genomic testing as a protytype. 
Later in Section~\ref{sec:eval}, we also present another use case of database queries for different input sizes.
Further potential use cases are discussed in brief in Appendix~\ref{sec:discuss}.

Single nucleotide polymorphism (SNP) is a common form of mutation in human DNA. 
Certain sets of SNPs determine the susceptibility of an individual to specific diseases. Analyzing an individual's set of SNPs  may reveal what kind of diseases a person may have.
More generally,
genomic data can uniquely identify a person, as it not only gives information about a person's association with diseases, but also about the individual's relatives~\cite{naveed2014privacy}.
Indeed, advancements in genomics research
have given rise to concerns about individual privacy and led to a number of related work in this space.
For instance, Canim et al.~\cite{canim2012secure} and Fisch et al.~\cite{fischiron} utilize tamper-resistant hardware to analyze/store health records.
Other works~\cite{wang2015efficient,baldi2011countering} investigate efficient, privacy-preserving analysis of health data.

While a number of different techniques have been proposed for privacy-preserving
genomic testing, ours is the first work to address this using one-time programs
grounded in secure hardware. Other than providing one-timeness, the proposed scheme also provides (i)~\textit{non-interactivity}, in which the user does not need to interact with the vendor during the protocol, and (ii)~\textit{pattern-hiding}, which ensures that the patterns used in vendor's test are kept private from the user. On the other hand, homomorphic encryption-based schemes~\cite{ayday2013privacy} lack non-interactivity and functional encryption-based schemes~\cite{naveed2014controlled}
lack non-interactivity and pattern-hiding.
We did not specifically implement these other techniques and compare our solution with them. However, from the performance results that are reported in the original papers,
we can argue that our proposed scheme provides comparable (if not better) efficiency compared to these techniques.

Our aim is to prevent the adversary (the client/Bob), who uses the
device for genomic testing,
from learning which positions of his genome are checked and how they are checked,
specifically for the genomic testing of the breast cancer (BRCA) gene. BRCA1 and BRCA2 are tumor suppressor genes.
If certain mutations are observed in these genes, the person will have an
increased probability of having breast and/or ovarian
cancer~\cite{walsh2010detection}. Hence, genomic testing for BRCA1 and BRCA2 mutations
is highly indicative of individuals' predisposition to develop breast and/or
ovarian cancer. 

We aim also to protect the privacy
of the vendor (the company/Alice) that provides the genomic testing and prevent
the case where the adversary extracts the test, learns how it works, and
consequently, tests other people without having to purchase the test. We aim to protect both the locations that are checked on the genome and the magnitude of the  risk factor corresponding to that position. Note that client's input is secure, as Bob is provided the device and he does not have to interact with Alice to perform the genomic test.

\vspace{-10pt}
\subsection{Genomic Test}
\label{sec:case_genotest}
\vspace{-5pt}
In order to perform our genomic testing, we obtained the SNPs related with BRCA1\footnote{Similarly, we can also list the SNPs for BRCA2 and determine the contribution of the observed SNPs to the total risk factor.} along with their risk factors from SNPedia~\cite{cariaso2010snpedia}, an open source wiki site that provides the list of these SNPs.
The SNPs that are observed on BRCA1 and their corresponding risk factors for breast cancer are listed in Appendix~\ref{sec:SNPTable}. 

We obtain genotype files of different people from the openSNP website~\cite{greshake2014opensnp}. The genotype files contain the extracted SNPs from a person's genome. At a high level, for each SNP of the patient that is linked to BRCA1, we add the corresponding risk factor to the overall risk.

The details of our genomic algorithm are shown in Appendix~\ref{sec:GeneticAlgo}. 
If a BRCA1-associated SNP is observed in the patient's SNP file, we check the allele combination and add the corresponding risk factor to the total amount. In order to prevent a malicious client from discovering which SNPs are checked, we check every line in the patient's SNP file. If an SNP related to breast cancer is not observed at a certain position, we add zero to the risk factor rather than skipping that SNP to prevent the client from inferring checked SNPs using side channels.

Let $i$ denote the reference number of an SNP and $s_i^j$ be the allele combination of SNP $i$ for individual $j$.  Also, $S_i$ and $C_i$ are two vectors keeping all observed allele combinations of SNP $i$ and the corresponding risk factors, respectively. Then, the equation to calculate the total risk factor for individual $j$ can be shown as $RF_j = \sum_{i} f(s_i^j)$ where

\[ f(s_i^j) = \begin{cases}
      C_i(\ell) & \textrm{ if $s_i^j = S_i(\ell)$ for $\ell=0,1,\ldots,|S_i|$} \\
      0 & \textrm{ otherwise} \\
   \end{cases} \]

For instance, for the SNP with ID $i=$ rs28897696, $S_i = \textless AA, AC\textgreater$ and $C_i = \textless7, 6\textgreater$. If the allele combination of SNP rs28897696 for individual $j$ corresponds to one of the elements in $S_i$, we add the corresponding value from $C_i$ to the total risk factor.

\vspace{-15pt}
\subsection{Construction for GC-Based}
\label{sec:case_gc}
The garbled circuit version of the genomic test presented in Section~\ref{sec:case_genotest}
is written as wire (.wir) code
accepted by the {\em Frigate} garbled circuit compiler.
The code follows the algorithm in Appendix~\ref{sec:GeneticAlgo},
adjusting overall risk factor upon comparing allele-pairs of matching SNPs
and explicitly adding zero when needed.

We choose Bob's input from AncestryDNA files available on
the openSNP website~\cite{greshake2014opensnp}. We perform preprocessing
on these to obtain a compact representation of the data
(specifics are available in Appendix~\ref{appendix:gcprep}).
Alice's input is hard-coded into the circuit at compile-time, by initializing an {\tt unsigned int} of vendor input size and assigning each bit's value using {\em Frigate}'s wire operator.

\vspace{-10pt}
\paragraph{Final input representation.}
Following the original design of {\em Battleship}, inputs are accepted as a single
string of hex digits (each 4 bits).
Each digit is treated separately, and input
is parsed byte-by-byte (e.g., $41_{16}$ is represented as
$10000010_{2}$).

We use 7 hex digits (28 unsigned bits) for the SNP reference number and
a single hex digit (4 unsigned bits) to represent the allele pair out of 16 possible
combinations of A/T/C/G. Alice's input contains
2 more hex digits (8 signed bits) for risk factor, supporting individual risk
factor values ranging from -128 to 127.
We keep risk factor a signed value, since some genetic mutations
lower the risk of disease. Although we did not observe any such mutations
pertaining to BRCA1, our representation gives extensibility to tests for other diseases.

\vspace{-10pt}
\paragraph{Output representation.}
The program outputs a signed 16-bit value,
allowing us to support
cumulative risk factor ranging from -32,768 to 32,767.\footnote{
This can easily be adjusted,
but is accompanied by substantial changes in the resulting circuit size.
For example,
an 11 GB circuit that outputs 16 bits grows to 18 GB by
doubling the output size to 32 bits.
We conservatively choose 16 bits for demonstration purposes,
but the output size may be reduced as appropriate.}

\vspace{-10pt}
\subsection{Construction for TXT-only}
\label{sec:case_txt}
\vspace{-5pt}
In TXT-only,
the genomic test logic of Section~\ref{sec:case_genotest} is ported in pure C
but largely keeps the representation used by the GC program (Section~\ref{sec:case_gc}).
Alice's input is in the form of
7 hex digits for the SNP ID, 1 hex digit for the allele pair and 2 digits for the risk factor. Bob's input is 2 digits shorter without the risk factor.

We pay special attention to minimizing exposure of Alice's input in RAM to defend against potential cold-boot attack.
We achieve this by processing one record at a time
performing all operations on and deleting it
before moving on to the next record.
We also seal each
record (10 bytes) into one sealed chunk (322 bytes), which
consumes more space. In each iteration, we unseal one of Alice's
records and compare with all of Bob's records. For certain laptops
and other computers with RAM soldered on the motherboard, this is
optional.



\vspace{-10pt}
\section{Performance evaluation}
\label{sec:eval}
\vspace{-5pt}
In this section, we evaluate the two OTP systems' performance/scalability, with varying client and vendor inputs,
and try to statistically verify the suitability of the two intuitive designs in different usage scenarios.
We perform our evaluation on a machine with a 3.50 GHz i7-4771 CPU,
Infineon TPM 1.2, 8 GB RAM,
320 GB primary hard-disk, additional 1 TB
hard-disk\footnote{We use a second disk to simulate what is shipped to the client (with all test data consolidated), separate from our primary disk for development.}
functioning as a one-time memory (dedicated to storing garbled circuit, and client and vendor input), running Ubuntu 14.04.5 LTS.
In one case, we required an alternate testing environment: a server-class machine with
a 40 core 2.20 GHz Intel Xeon CPU and 128 GB of RAM.\footnote{
Another option would have been to upgrade the memory of the initial evaluation machine,
but we chose to forgo this, as a test run on the server-class machine revealed
that upwards of 60 GB would be required (not supportable by the motherboard).}
Details specific to the setup of our genomic testing were previously given in Section~\ref{sec:case}. \looseness=-1

We perform experiments to determine the effects of varying either client or vendor input size.
Based on the case study, the vendor has 880 bits
and the client has 22.4M bits of input, so we use 224 and 880 as the base numbers for our evaluation.
We multiply by multiples of 10 to show the effect of order-of-magnitude changes on inputs.
We start with 224 for client and 880 for vendor inputs.
When
varying client input, we fix vendor input at 880 bits. When varying vendor input,
we fix client input at 224K bits.

\vspace{-10pt}
\subsection{Benchmarking TXT-only}
\label{sec:eval-txtonly}
\vspace{-5pt}

\def \hfillx {\hspace*{-\textwidth} \hfill}

\subhead{Varying client input}
Table~\ref{tbl:txtonlyc} shows the timing results for TXT-only provisioning and execution with
fixed vendor input and varying client input size.
During provisioning, only the vendor input is sealed, so the provisioning time
is constant in all cases. As client input size increases, so does execution time,
but moderately. Performance is insensitive to client input size up through the 224K case. Even for the largest (22M) test case, increasing the client input size by two orders of magnitude results only in a
slowdown by a factor of 3.5x.

\subhead{Varying vendor input}
Table~\ref{tbl:txtonlyv} shows the timing results with fixed
client input and varying vendor input size. Although we only tested against three configurations, we see an order-of-magnitude increase in vendor input size is
accompanied by an order-of-magnitude increase in both provisioning and execution times.

\begin{table}[t]
	\begin{minipage}{0.5\textwidth}
			\scriptsize
			\centering
			\begin{tabular}{|c||c|c|c|c|}
				\hline
				\begin{tabular}[c]{@{}c@{}}Client Input (bits)\end{tabular} & Prov. (ms) & Exec. (ms) \\ \hline
				\hline
				224 & 5640.17 & 9394.58 \\ \hline
				2K & 5640.17 & 9393.88 \\ \hline
				22K & 5640.17 & 9388.27 \\ \hline
				224K & 5640.17 & 9426.56 \\ \hline
				2M & 5640.17 & 11078.19 \\ \hline
				22M & 5640.17 & 33427.50 \\ \hline
			\end{tabular}
            \vspace{5pt}
			\caption{TXT-only results with vendor input fixed at 880 bits and varying client input size, averaged over 10 runs. Prov./Exec. refers to the provisioning mode and execution mode respectively.} 
			\label{tbl:txtonlyc}
	\end{minipage}
	\hfillx
	\begin{minipage}{0.45\textwidth}
			\scriptsize
			\centering
			\begin{tabular}{|c||c|c|c|c|}
				\hline
				\begin{tabular}[c]{@{}c@{}}Vendor Input (bits)\end{tabular} & Prov. (ms) & Exec. (ms) \\ \hline
				880   & 5640.17   & 9426.56 \\ \hline
				8800  & 53515.75  & 92551.43 \\ \hline
				88000 & 527026.89 & 921338.53 \\ \hline
			\end{tabular}
            \vspace{5pt}
			\caption{TXT-only results with client input fixed at 224k bits and varying vendor input size,
				averaged over 10 runs.
				Performance of TXT-only is linear and time taken is proportional to vendor input size.}
			\label{tbl:txtonlyv}

	\end{minipage}
\vspace{-24pt}
\end{table}

\vspace{-10pt}
\subsection{Benchmarking GC-based}
\label{sec:eval-gc}
\vspace{-5pt}
We use the same experimental setup as used in TXT-only, but with additional time taken by the GC portion. 
Vendor and client each incur runtime costs from a GC ({\tt gen}/{\tt evl}) and
a sealing-based (Prov./Sel.) phase.

\subhead{Varying vendor input}
We are interested in whether~\emph{GC-based} is less sensitive to the size of Alice's
input than~\emph{TXT-only}; see Table~\ref{tbl:gcbasedv}.
Since provisioning (Prov.) involves sealing a constant number of key pairs, and
selection (Sel.) is dependent on the unsealing of these key pairs to output one key from each,
there is no change. Both {\em Battleship} {\tt gen} and {\tt evl} mode timing
is largely invariant, as well.
Whereas System 1 performance was linearly dependent on vendor input size, we observe
that~\emph{GC-based} (System 2) is indeed not sensitive to vendor input.

\begin{wraptable}{r}{0.55\textwidth}
\vspace{-20pt}
		\scriptsize
		\centering
		\begin{tabular}{|c||c|c|c|c|}
			\hline
			\begin{tabular}[c]{@{}c@{}}Vendor \\ Input (bits)\end{tabular} & {\tt gen} (ms) & Prov. (ms) & Sel. (ms) & {\tt evl} (ms) \\ \hline
			\hline
			880 & 2323.7 & 4244.03   & 2508.73  & 31815.4\\ \hline
			8800  & 3198.7 & 4244.03  & 2508.73 & 32200.4 \\ \hline
			88000  & 3286.9 & 4244.03 & 2508.73 & 32000.9 \\ \hline
		\end{tabular}
        \vspace{-5pt}
		\caption{GC-based results with client input fixed at 224k bits, varying vendor input size, and
			encryption of keys by a sealed master key,
			averaged over 10 runs.}
		\label{tbl:gcbasedv}
		\vspace{-20pt}	
\end{wraptable}

\subhead{Varying client input}
For completeness, we also examine the effects of varying client input size on runtime;
see Table~\ref{tbl:gcbasedc}. Prov.\ and Sel.\ stages are both slow
as client input size increases, since more key pairs must be sealed/unsealed. {\tt gen}
and {\tt evl} times are also affected by an increase in client input bits. Most notably,
{\tt evl} demonstrates a near order-of-magnitude slowdown from the 224K case to the 2M case, and the slowdown trend continues into the 22M case (despite using the better-provisioned
machine to evaluate the 22M case).
We indeed find that TXT-only OTP is complemented by GC-based OTP, where
performance is sensitive to client input.

\begin{table}[t]
    \begin{minipage}{0.57\textwidth}

		\scriptsize
		\centering
		\begin{tabular}{|c||c|c|c|c|}
			\hline
			\begin{tabular}[c]{@{}c@{}}Client\\Input (bits)\end{tabular} & {\tt gen} (ms) & Prov. (ms) & Sel. (ms) & {\tt evl} (ms) \\ \hline
			\hline
			224  & 1503.7 & 843.64    & 600.55 & 1350.8 \\ \hline
			2K   & 1318.9 & 906.70    & 688.62 & 1631.8 \\ \hline
			22K  & 1659.7 & 991.91    & 724.24  & 3643.7 \\ \hline
			224K & 2323.7 & 4244.03   & 2508.73  & 31815.4 \\ \hline
			2M   & 16842.8 & 33934.54  & 19188.31 & 305362.8 \\ \hline
			22M  & 148387.9* & 346606.87 & 283704.57 & 3108271* \\ \hline
		\end{tabular}
        \vspace{3pt}
		\caption{GC-based results with vendor input fixed at 880 bits, varying client input size, and
			encryption of keys by a sealed master key,
			averaged over 10 runs.
			Provisioning- and execution-mode times were measured separately.
            *s indicate tests run in an alternate environment, due
to insufficient memory on our primary testing setup.}
		\label{tbl:gcbasedc}
    \end{minipage}
    \hfillx
	\begin{minipage}{0.39\textwidth}
		\centering

        \scriptsize
		\begin{center}
			\begin{tabular}{|c|c|c|}
				\hline
				OTP Type & Mode & Timing (ms) \\ \hline
				\hline
				\multirow{2}{*}{TXT-only} & Prov. & 5640.17 \\ \cline{2-3}
				& Exec. & 33427.50 \\ \hline \hline
				\multirow{4}{*}{GC-based} & {\tt gen} & 148387.9* \\ \cline{2-3}
				& {Prov.} & 346606.87 \\ \cline{2-3}
				& {Sel.} & 283704.57 \\ \cline{2-3}
				& {\tt evl} & 3108271* \\
				\hline
			\end{tabular}
		\end{center}
        \vspace{-5pt}
		\caption{Performance for TXT-only and GC-based OTP
			implementations of the BRCA1 genomic test,
			averaged over 10 runs. Vendor input is 880
			bits. Client input is 22,447,296 bits.
            *s indicate tests run in an alternate environment, due
to insufficient memory on our primary testing setup.
		}
		\label{tbl:caseresults}
	\end{minipage}
\vspace{-26.5pt}
\end{table}

\vspace{-10pt}
\subsection{Analysis}
\label{sec:results}
\vspace{-5pt}

Onto our real-world genomic test (among other padded data sets for the evaluation purpose),
Alice's input comprises the 22 SNPs associated
with BRCA1, presented in Table~\ref{Table:SNPsBRCA} of Appendix~\ref{sec:SNPTable}.
Each SNP entry takes up
40 bits, so Alice's input takes up 880 bits.
Bob's input comprises the 701,478 SNPs drawn from his AncestryDNA file, each of which
is represented with 32 bits, adding up to a total size of 22,447,296 bits.
This genomic test corresponds to our earlier experiment with vendor input size of 880 bits and client input size of 22M bits.

\begin{wraptable}{r}{0.37\textwidth}
\vspace{-30pt}
		\centering
            \scriptsize	
			\begin{center}
				\begin{tabular}{|c||c|}
					\hline
					\begin{tabular}[c]{@{}c@{}}Small Vendor +\\Small Client\end{tabular} &
					\begin{tabular}[c]{@{}c@{}}Small Vendor +\\Large Client\end{tabular} \\
					& \\
					{\bf TXT-only} & {\bf TXT-only} \\ \hline \hline
					\begin{tabular}[c]{@{}c@{}}Large Vendor +\\Small Client\end{tabular} &
					\begin{tabular}[c]{@{}c@{}}Large Vendor +\\Large Client\end{tabular} \\
					& \\
					{\bf GC-based} & {\bf TXT-only} \\
					\hline
					
					\hline
				\end{tabular}
			\end{center}
            \vspace{-15pt}
			\caption{Depending on the input sizes of vendor and client,
				one system may be preferred to the other. GC-based OTP is favorable
				when large vendor input is paired with small client input; TXT-only OTP otherwise.}
			\label{tbl:quadrants}
		\vspace{-30pt}

\end{wraptable}

Table~\ref{tbl:caseresults} puts together the results for both OTP systems.
Even at first glance, we see that TXT-only OTP vastly outperforms the GC-based OTP. 
Provisioning is two orders of magnitude slower in GC-based OTP,
and trusted selection itself is an order of magnitude slower than the entire
execution mode of TXT-only OTP.
{\tt gen} and {\tt evl} further introduce a performance hit to GC-based OTP
(again, despite the fact that we evaluated this case on a 
better-provisioned machine).
TXT-only is the superior option for our genomic application.

\vspace{-5pt}
\paragraph{Choosing one OTP.}
We already saw in Section~\ref{sec:eval-txtonly} that TXT-only OTP is less sensitive to
client input, whereas we saw in Section~\ref{sec:eval-gc} that GC-based OTP is less
sensitive to vendor input. We illustrate the four cases in Table~\ref{tbl:quadrants} in
four quadrants.

In this specific use-case of genomic testing, we are in the upper-right quadrant and thus the TXT-only OTP dominates. However, other use cases, like the database querying examples in Appendix~\ref{sec:discuss}, occupy the lower-left quadrant, in which case GC-based will outperform the TXT-only OTP. What should we do if both inputs are of similar size (i.e., equally ``small'' or ``large'')? A safe bet is to stick with the TXT-only OTP. Even though GC technology continues to improve, garbled circuits will always be less efficient than running the code natively. \looseness=-1

\subsection{Another use case: database queries.}
To give an example where the vendor input can be significantly large, we may consider another potential and feasible application of our proposed OTP designs, where GC-based can outperform TXT-only.
It is also in a medical setting where the protocol is between two parties, namely a company that owns a database consisting of patient data and a research center that wants to utilize patient data. The patient data held at the company contains both phenotypical and genotypical properties. The research center wants to perform a test to determine the relationship of a certain mutation (e.g., a SNP) with a given phenotype. There may be three approaches for this scenario:

\begin{enumerate}
  \item \textbf{Private information retrieval~\cite{chor1995private}:} PIR allows a user to retrieve data from a database without revealing what is retrieved. Moreover, the user also does not learn about the rest of the data in the database (i.e., symmetric PIR~\cite{saint2005java}). However, it does not let the user compute over the database (such as calculating the relationship of a certain genetic variant with a phenotype among the people in the database).
  \item \textbf{Database is public, query is private:} The company can keep its database public and the research center can query the database as much as it wants. However, with this approach the privacy of the database is not preserved. Moreover, there is no limit to the queries that the research center does.

  As an alternative to this, database may be kept encrypted and the research center can run its queries on the encrypted database (e.g., homomorphic encryption). The result of the query would then be decrypted by the data owner at the end of the computation~\cite{kantarcioglu2008cryptographic}. However, this scheme introduces high computational overhead.

  \item \textbf{Database is not public, query is exposed:} In this approach, the company keeps its database secret and the research center sends the query to the company. This time the query of the research center is revealed to the company and the privacy of the research center is compromised.

\end{enumerate}

In the case of GC-based, the company stores its database into the device (in the form of garbled circuit) and the research center purchases the device to run its query (in TXT) on it. This system enables both parties' privacy. The device does not leak any information about the database and also the company does not learn about the query of the research center, as the research center purchases the device and gives the query as an input to it. In order to determine the relationship of a certain mutation to a phenotype, chi-squared test can be used to determine the p-value, that helps the research center to determine whether a mutation has a significant relation to a phenotype. We leave this to future work.



\vspace{-10pt}
\section{Security analysis}
\label{sec:security}
\vspace{-5pt}

\subhead{a) Replay attacks} The adversary may try to trick the OTP into executing multiple times by replaying a previous state, even without compromising the TEE, or the one-time logic therein. The secrets (e.g., MK) only have per-deployment freshness (fixed at Alice's site). Nevertheless, in our implementation, the TPM NVRAM indices where the one-timeness flag and MK are stored are configured with PCR-bound protection, i.e., outside the correct environment, they are even inaccessible for read/write, let alone to replay.

\subhead{b) Memory side-channel attacks} Despite the hardware-aided protection from TEE, sensitive plaintext data must be exposed at certain points. For instance, MK is needed for encrypting/decrypting key pairs, and the key pairs when being selected must also be in plaintext. Software memory attacks~\cite{foreshadow,meltdown,spectre}
do not apply to our OTP systems, as the selected TEE (TXT) is exclusive. In our design, the code running in TEE does not even involve an OS, driver, hypervisor, or any software run-time. There are generally two categories of physical memory attacks: non-destructive ones that can be repeated (e.g., DMA attacks~\cite{dma-thunderbolt}); and the destructive (only one attempt) physical cold-boot attack~\cite{cold-boot}.  All I/O access (especially DMA) is disabled for the TEE-protected regions and thus DMA attacks no longer pose a threat.

The effective cold-boot attack requires that the RAM modules are swappable and plaintext content is in RAM. 
For certain laptops or barebone computers~\cite{ram}, their RAM is soldered on the motherboard and completely unmountable (and thus immune). To ensure warm-boot attacks~\cite{warm-boot} (e.g., reading RAM content on the same computer by rebooting it with a USB stick) are also prevented, we can set the Memory Overwrite Request (MOR) bit to signal the UEFI/BIOS to wipe RAM on the next reboot before loading any system (cf. the official TCG mitigation~\cite{tcgwarm}).
We do take into account the regular desktops/laptops vulnerable to the cold-boot attack:
For small-sized secrets like MK, existing solutions~\cite{tresor,copker,pixelvault,amnesia} can be used,
where CPU/GPU registers or cache memory are used to store secrets. For larger secrets, like the key pairs/vendor input, we perform block-wise processing so that at any time during the execution, only a very small fraction is exposed. Also, as cold-boot attack is destructive, the adversary will not learn enough to reveal the algorithm or reuse the key pairs. At least, the vendor can always choose computers with soldered-down RAM.

\subhead{c) Attack cost} 
Bob may try to infer the protected function and vendor inputs by trying different inputs in multiple instances. This attack may incur a high cost as Bob will need to order the OTP from Alice several times.
This is a limitation of any offline OTP solution, which can only guarantee one query per box.

\subhead{d) Cryptographic attacks} The security of one-time programs (and garbled circuits) is proven in the original paper~\cite{goldwasserOTP} (updated after caveat~\cite{BHR12}), so we do not repeat the proofs here.

\subhead{e) Clonability} Silicon attacks on TPM can reveal secrets
(including the Endorsement Key), but chip imaging/decapping requires high-tech
equipment. Thus, cloning a TPM or extracting an original TPM's identity/data to populate
a virtual TPM (vTPM) is considered unfeasible. Sealing achieves platform-state-binding without
attestation, so non-genuine environments (including vTPM) will fail to unseal. Refer to Appendix~\ref{sec:relay-smm-attacks} for TPM relay attack and  SMM attacks. Furthermore, there has been a recent software attack~\cite{baddream} that resets and forges PCR values during S3 processing exploiting a TPM 2.0 flaw (SRTM) and a software bug in tboot (DRTM). They (allegedly patched) do not pose a threat to our OTP design, as neither SRTM nor any OS software (e.g., Linux) is involved, not to mention our OTP does not support/involve any power management.

\subhead{f) Input credibility/correctness for genomic tests} Genomic tests should be run with the user's consent and an attacker shouldn't be able to run tests with fake genomes
to infer the test. We can use digital signatures to provide credibility, and use biometric attributes to ensure ownership. Another related concern is \emph{inference attacks:} Genetic disorders
may be highly correlated with each other; e.g., the SNP with ID rs429358 has influence on the risk of having both Alzheimer's disease and heart disease~\cite{snpcorrelation}.
Thus the result of a single genomic test can still give information about other diseases. Moreover, circuits should be designed with
the same circuit depth to prevent Bob from inferring Alice's input.
Additionally, unless a (projective) garbling scheme that realizes an OTP is designed to provide
adaptive privacy, it opens up room for \emph{adaptive attacks}~\cite{BHR12}.\footnote{For certain classes of circuits, Jafargholi and Wichs~\cite{JW16} claim that garbled circuits are adaptively secure without further 	modification, with security loss tied to pebble complexity of the circuit.} Solutions include: allowing the adversary to decrypt the circuit
but not learn the output of the circuit until all keys have been chosen~\cite{goldwasserOTP}; encrypting the circuit using either a one-time-pad or random-oracle-based encryption and revealing the decryption key together with the garbled input in the online phase~\cite{BHR12}; and placing a ``holdoff" gate into each output wire that cannot be evaluated until all keys are learned~\cite{fpga-otp}.



\section{Concluding Remarks}
\label{sec:conc}

Until now, one-time programs have been theoretical or required highly customized/expensive hardware. 
We shift away from crypto-intensive approaches to the emerging but time-tested trusted 
computing technologies, for a practical and affordable realization of OTPs.
With our proposed techniques, which we will release publicly, anyone can build a one-time program 
today with off-the-shelf devices that will execute quickly at a moderate cost. The cost of our 
proposed hardware-based solution for a single genomic test 
can be further diluted by extension to support multiple tests
and multiple clients on a single device (which our current construction already does).
The general methodology we provide can be adapted to other trusted execution environments to satisfy
various application scenarios and optimize the performance/suitability for existing applications.



\appendix

\newpage
\noindent
The appendices are organized as follows:
\begin{itemize}
\item Appendix~\ref{appendix:bg} provides additional background helpful for understanding on one-time programs, garbled circuits, and one-time memories;
\item Appendix~\ref{app:adaptive} addresses an adaptive security attack on OTP systems;
\item Appendix~\ref{sec:frigmods} describes in detail modifications we make to {\em Battleship}; 
\item Appendix~\ref{appendix:gcprep} presents preprocessing steps for our case study application;
\item Appendix~\ref{sec:discuss} provides additional one-time program use cases; 
\item Appendix~\ref{sec:SNPTable} lists the SNPs associated with BRCA1;
\item Appendix~\ref{sec:GeneticAlgo} gives our genomic algorithm;
\item Appendix~\ref{appendix:port} comments on porting efforts required for OTP; and
\item Appendix~\ref{sec:relay-smm-attacks} discusses SMM and TPM relay attacks.
\end{itemize}

\section{Additional Background}
\label{appendix:bg}

\subsection{One Time Programs}
\label{appendix:otp}

A one-time program (OTP), as introduced by Goldwasser \etal~\cite{goldwasserOTP}, is an implementation of a deterministic function which is provided by Alice to Bob. We describe it here with less generality than it was presented originally (see Section \ref{sec:threat} for a reconciliation of both approaches). Consider the implementation as containing the function itself (unprotected) and Alice's input to the function (cryptographically protected). Bob can choose a single input and evaluate the function (with Alice's input) on it. With the output, he may be able to infer something about Alice's input (depending on the exact function), but he cannot infer anything about her input beyond this. Since Bob is operating the device autonomously from Alice, his input is unconditionally private from Alice. The core requirement of OTP is that while Bob is able to receive the evaluation on a single input of his choosing, he is unable to obtain an evaluation of any other input. The mechanism to enforce this is the topic of this paper.

The term `one-time program' is a slight misnomer. One might equate it with a form of copy protection or digital rights management (DRM). It is worth illustrating the difference with a simple example. Consider a DRM scenario: Alice providing Bob with a media player (the function) and a movie (Alice's input). The movie will play if Bob inputs a correct access code. In this case, the stream of the movie is the output of the function; once Bob learns the output, he can replay it as many times as he wants. Therefore, this is not a valid application of OTP. Instead, consider the following: Alice provides Bob with a game of Go (the function) programmed with the latest in artificial intelligence (Alice's input). Bob's moves are his input to the function. He can `replay' the game with the exact same moves (resulting in the exact same game and outcome) as many times as he wants (so it is not strictly `one-time'), however as soon as he deviates with a different move, the program will not continue playing. In this sense, he can only play `once,' limited to a single sequence of moves.

OTPs can be realized in a straightforward way with trusted execution. From here, we will describe the alternative approach~\cite{goldwasserOTP} of realizing one-time programs via a simpler primitive called one-time memory (OTM), and composing OTM with garbled circuits.

\paragraph{Garbled Circuits.}
\label{appendix:gc}

Garbled circuits (GC) were first proposed by Yao~\cite{yao_gc} as a technique for
achieving secure multiparty computation by at least two parties, a
{\em generator} (Alice) and an {\em evaluator} (Bob). A program is first
converted into its Boolean circuit representation.
For each of the $i$ wires in the circuit, Alice chooses encryption keys
$k^0_i$ and $k^1_i$. Each gate of the circuit takes on the form of a truth table,
and entries of the truth table are permuted to further conceal whether any
particular entry holds a 0- or 1-value.
The keys received on each input wire unlocks a single entry of the truth table,
itself a key that is released on the output wire and fed into the next gate.
Bob receives the garbled circuit from Alice, together with
Alice's garbled inputs.
Bob garbles its own inputs through oblivious transfer (OT) with Alice.
During evaluation, an output key is iteratively unlocked, or decrypted, from
each of the garbled gates until arriving at the final output,
which is revealed to all participants.

\paragraph{One-Time Memory.}
\label{appendix:otm}

In summary, one-time programs extend garbled circuits where the oblivious transfer phase is replaced with a special purpose physical device called one-time memory (OTM). The protocol proceeds as in garbled circuits with Bob given the circuit, encoded with Alice's input under encryption. Instead of interacting with Alice to learn the keys that correspond to his input, Alice provisions a device with all keys on it. However when Bob reads a key off the device (say for input bit 0) the corresponding key (for input bit 1) is erased. The end result is the same as oblivious transfer: Bob receives exactly one key for each input bit while not learning the other key, whereas Alice does not learn which keys Bob selected. The main difference is that key-selection by OTM is non-interactive, meaning Alice can be completely offline.

\subsection{Trusted execution environments} \label{sec:tee}

Based on the desired system properties defined in Section~\ref{sec:threat}, we now discuss the requirements a candidate TEE should satisfy. 
\begin{itemize}
\item \textbf{R1}: Isolated execution with integrity. This corresponds to Property 3, so that the logic is properly enforced and no additional runs are allowed. Most TEEs have this fundamental feature.
\item \textbf{R2}: Non-volatile secure storage (formally termed Secure Element). Particular to OTP,  intuitively a non-volatile flag is needed to record if the program has been run (or how many times). This is required for all stateful programs.
\item \textbf{R3}: Sealing (machine state binding). For Property 1, the non-run-time secrecy of $a$ needs the capability to bind it to the exact desired machine state, and only under this state can it be retrieved. Such capability is usually called sealing in most TEEs.
\item \textbf{R4}: No software side-channels. To ensure run-time secrecy, there should be no way for other code (if any) on the same device to learn the secret (or other protected data). Exclusive TEEs naturally satsfy this requirement.
\item \textbf{R5}: No physical side-channels. There should be no physical side-channels, which specifically refers to either (DMA-related) memory attacks and the cold-boot attack.
\end{itemize}
 Even if not all the requirements (\textbf{R1} - \textbf{R5}) can be satisfied by a particular TEE, we would like to see which is best-positioned to realize OTP and what additional steps can be taken to compensate for those that are missing.

Trusted computing (where TEEs belong) already has a history of more than a decade (cf. 
an earlier endeavor of Texas Instrument M-Shield~\cite{mshield} on OMAP). TEEs are usually 
architecture-shipped, with a primary focus on securing processor execution.
They can be categorized as one of the following: 
\vspace{-5pt}
\begin{compactlistn}
\item Exclusive. Exemplified by Intel TXT, this type
of TEE suspends all other operations on the processor and owns all resources
before it exits. The advantage is less attack vectors exposed.
\item Concurrent. Represented
by Intel SGX and ARM TrustZone, this type creates secure enclaves or worlds that exist alongside
other processes. There might be multiple instances at the same time. These are more
suitable for application-level logic. 
\end{compactlistn}
\vspace{-5pt}
We now present a few of the typical TEE
options in the context of OTP, and discuss their suitability for matching each of our stated requirements.
All TEEs satisfy \textbf{R1}, without need for explicit deliberation.

\bfhead{Intel TXT~\cite{TXT} and AMD SVM~\cite{amdsvm}.}
TXT and SVM are simply counterparts on their respective vendor's platform, with nearly the same properties
(slight differences). 
They are exclusive by nature and rely on a security chip called TPM (Trusted Platform Module),
corresponding to \textbf{R2}.
When the secure session is started anytime, TXT/SVM measures the loaded binaries and stores
the results in TPM. Two primitives are important: 1) Measured launch. TXT/SVM can compare
the measurements with the ``good'' values in TPM and aborts execution if mismatch occurs.
2) Sealing (platform binding, satisfying \textbf{R3}). Sealed data can only be accessed in the intact, 
genuine program and correct platform. Their exclusiveness naturally meets \textbf{R4}, as no other
code can be run simultaneously. As desktop processors, detachable RAM modules are inevitable,
so the cold-boot attack fails \textbf{R5}. We will discuss a workaround in Section~\ref{sec:txtonly}.

\bfhead{ARM TrustZone~\cite{trustzone}.}
TrustZone introduces the notion of secure world and normal world. The secure world
coexists with the normal world, with everything (including I/O) separated. The two
can communicate through a special monitor. This leaves it questionable for \textbf{R4},
as there might be potential side-channel attacks from code running in the normal world.
Since it is coupled with the ARM architecture,
we can use it on mobile platforms or a dedicated device other than a desktop.
This intrinsically satisfies \textbf{R5} as it should be difficult (if not impossible) to 
physically extract RAM secrets, e.g., by probing or detaching memory modules.
TrustZone also supports sealing satisfying \textbf{R3}.
An obvious advantage of TrustZone is its secure peripheral communication
(enabled by the AMBA3 AXI to APB Bridge).
For example, if a small region of the screen
is allocated to the secure world, user input there cannot be intercepted by
the normal world OS. However, in our OTP, we have no need to involve a
regular full-blown OS. Moreover, one of its disadvantages is that the essential
secure element (where secrets are stored, like TPM) is not standardized
and always vendor-specific~\cite{op-tee}, thus failing \textbf{R2}.
This means for any OTP we develop, we have to collaborate closely with
the device manufacturer, whereas for TXT, we can buy COTS devices.
Nevertheless, if such collaboration became possible for a specific organization, 
making use of TrustZone on mobile platforms can significantly lower the cost
(from approximately \$500 to a few dollars) per device.

\bfhead{Intel SGX~\cite{sgx}.} More recent than TXT, SGX (Software Guard Extensions) can also be utilized to achieve one-timeness. 
Intel SGX provides finer-grained isolated environment (measurement-based like TXT) where individual 
secured apps (called \emph{enclaves}) coexist with the untrusted operating system (thus failing \textbf{R4}). 
The integrity of the 
program logic (e.g., refusing to run a second time) is guaranteed by the measurement of enclaves before 
loading. However, what was missing has been a secure persistent storage for the flag (to ensure one-timeness) 
and Alice's input (SGX did not use TPM in the first place); without secure storage, Bob can simply make a copy of both 
before execution/evaluation, hence defeating one-timeness. To bring back freshness with SGX-sealed data, 
Intel recently added support for non-volatile on-chip monotonic counters (similar to TPM, stored in the SPI flash chip), see ROTE~\cite{rollback}. 
Therefore, SGX-sealing the flag and key pairs with replay attack resistance is feasible now (\textbf{R3}).
We consider its \textbf{R2} to be partial, as there is not dedicated general-purpose secure storage like TPM. 
With respect to \textbf{R5}, SGX enclaves' memory is always encrypted outside the CPU, thus immune to the cold-boot attack.\\

\noindent
There have also been other TEEs around that are not discussed here, for the reason that either they are
less-used or obsoleted (e.g., M-Shield~\cite{mshield}) or no sufficient public information 
is available to support
development (e.g., Apple Secure Enclave Co-processor~\cite{ios}). We decide to implement our engineering 
prototype with Intel TXT with the following considerations (compared with SGX). Note that since
TrustZone requires vendor collaboration, we skip it for now.

\begin{compactlistn}
\item \textbf{Fewer known flaws.} TXT has been time-tested and known flaws are already stable public information (see Section~\ref{sec:security}).
For SGX, there have been multiple reports regarding various side-channel attacks mounted by malicious/compromised OS or even peer apps~\cite{xu2015controlled,malwaregx}. What is worse, Intel admits it as a known flaw that will remain, leaving the closing of side-channels as a responsibility of enclave developers~\cite{stancesc}.
In other words, side-channels are explicitly out of the thread model of SGX.
Such a flaw allows potential multiple or even unlimited number of executions 
of the protected program, which Bob is motivated to do. On the other hand,
although TXT used to have a few system/hardware-level flaws~\cite{attack-txt,another-txt} (as no other software can coexist), there are no recent such
reports, and previous ones have been patched or not-applicable any more with newer CPU versions. Note that certain attacks based on SMM (System Management Mode) have
also been targeting TXT, but does not pose as much threat here, as explained 
in Section~\ref{sec:relay-smm-attacks}.
\item \textbf{Meltdown~\cite{meltdown}/Spectre~\cite{spectre}/Foreshadow~\cite{foreshadow}.} The lately identified flaws in modern processors make side-channel attacks potentially ubiquitous, due to the fact that out-of-order execution is a common feature of modern architectures. What make it worse is the Foreshadow attack specifically targeting SGX (L1 Terminal Fault). Even after a microcode patch, the trustworthiness is still gone as there is no guarantee that the unique per-CPU key was not exposed before the patch. All such intrinsic side-channel attacks stem from multitasking (co-existing programs). Therefore,the \emph{exclusive} trusted environment (where no other OS/entities/processes exist) is more preferable in achieving one-timeness, which is the case for TXT.
\item \textbf{Dedicated environment.} SGX is positioned differently than TXT and does not replace it, in the sense that the former allows multiple user-space instances for cloud applications, whose attestation requires contacting Intel's IAS server each time. In contrast, TXT is a dedicated environment, with reduced attack vectors, that also allows local attestation.
\end{compactlistn}

\section{Adaptive Attacks on OTPs}
\label{app:adaptive}

A subtle security issue arises when comparing the views in the GC-based variant: $\tuple{\mathsf{OTP},f(a,b),b}$ and $\tuple{f(a,b),b}$. In the former case, Bob specifies $b$ bit-by-bit and starts to learn information about $f(a,b)$ before choosing all bits of $b$. Bob can thus adaptively choose the next bits of $b$ based on his observations about the path through the circuit, and the number of possible paths is generally exponential in the number of gates. Thus, in a formal simulation-based proof, the simulator cannot add a convincing $\mathsf{OTP}$ to the second view without knowing all of $b$ a priori.\footnote{This problem does not arise in garbled circuits since the oblivious transfer is completed prior to providing the circuit.} Goldwasser \etal\ add an all-or-nothing transformation to the output of $f(a,b)$: the output is masked by random bits that are only fully revealed once all of $b$ is selected.\footnote{This fixes the issue because the simulator can equivocate on the final masking value to program the random value, sitting wherever the circuit ends up, with the correct output value for the now-known input.} In the TXT-only variant, the output is provided at the end as enforced by the execution of the system.

\section{The Frigate GC compiler}
\label{sec:frigmods}

\paragraph{Frigate.}{\em Frigate}~\cite{frigate} is 
a modern Boolean circuit compiler that
outperforms several other garbled-circuit
compilers (e.g., PCF~\cite{pcf}, Kreuter \etal~\cite{kss},
CBMC~\cite{cbmc}) by orders of magnitude.
{\em Frigate} is also extensively validated
and found to produce correct and functioning circuits where other compilers
fail~\cite{frigate}.
For these reasons, we decide to use {\em Frigate} for implementing
the garbled circuit components of our GC-based OTP.\footnote{Although we choose to go with {\em Frigate}, it is possible to instantiate our
OTP system with other garbled circuit compilers.}

\paragraph{Battleship.}{\em Battleship}, developed by the same group, 
separates out the interpreter and execution functionalities 
of {\em Frigate}.
Battleship reads in and interprets the
circuit file produced by {\em Frigate}.
{\em Battleship} is originally designed to be run interactively
by at least two parties,
a generator and an evaluator.
The generator is able to independently
garble its own inputs, whereas the evaluator
depends on OT to garble its inputs.
At each gate of the garbled circuit, a single value is decrypted from the
associated truth table containing encrypted entries. 
Garbled gates are iteratively decrypted until arriving at the final output,
which is released to either party.
Output need not be the same for both parties.

To make {\em Battleship} support one-time programs, we do the following:
\begin{compactlist}
\item Split execution in {\em Battleship} into two standalone phases. In the first phase,
a fresh set of random keys (0- and 1-keys for encoding each bit of the client's input) is
generated and written out to a file. The key pairs contained in this file will be
used during TXT provisioning, after which the file is discarded.
The second phase reads in another file containing the subset
of keys chosen (during trusted selection) according to
the client's input bits and performs evaluation of the circuit. 
{\em Battleship} did not originally require these file operations since
inputs were garbled and immediately usable
without needing to interrupt the system, while we rely on Intel TXT.
\item Remove the oblivious-transfer step. In our setting, vendor and client do not perform
interactive computation in real time. Instead, the client receives the garbled representation
of its input during the trusted selection process inside Intel TXT. The client's input chosen
in this way is not exposed to the vendor, who no longer
has access to the system after sending it to the client.
\item Remove dependency on the full set of 0- and 1- keys in the second phase.
In the original {\em Battleship} design, generator and
evaluator would be separate parties, so the full set of keys were not visible to the evaluator
even though it remained available to the generator over the course of evaluation.
In our setting, both generator and evaluator functions run on the same machine,
so it is imperative that we make the full key set unavailable.
We achieve this by instead supplying both the chosen subset of keys and
the raw binary input of the client into the second phase of our modified {\em Battleship}.
In this way, individual keys can be identified as either 0- or 1-keys without
needing to examine the full key set.
\end{compactlist}

\section{GC-based Case Study Setup}
\label{appendix:gcprep}

\paragraph{Client input.} To arrive at a compact representation of Bob's input, we employ a simple
bash shell script to:
\begin{itemize}
    \item Remove unused chromosome and position fields.
    \item Remove comment lines at the start of file and line containing field headings.
    \item Remove the ``rs'' prefix from each SNP reference \#.
    \item Remove all spaces between fields and line breaks between entries,
making the entire input one line.
    \item Convert SNP reference numbers into hexadecimal format,
and zero-pad the result to length 7 (hex format allows us to
reduce 4 keys per entry for a more efficient representation).
    \item Merge Allele 1 and Allele 2 fields, and assign a 1-digit
hex value to each possible allele pair.
\end{itemize}
The removal of all spaces and line breaks caters to the original
{\em Battleship} design,
which expects inputs to be read in as a single line.
It is especially important that reference numbers be padded with zeroes
(e.g., 0x3DE2 (15842), becomes 0x0003DE2),
given that we merge all inputs into a single line, so
entries can be parsed using fixed indices.
7 hex digits is sufficient to support all reference numbers,
which have at most 8 decimal digits.

\paragraph{Vendor input.}
If a particular SNP has more than one ($allele~pair$, $risk~factor$) mapping,
then each of these is treated as a separate input (with SNP reference number repeated).
Although this leads to increased circuit size,
specifying Alice's input in this manner is necessary in order to avoid
subtle timing disparities which may leak information about the test being performed.
The alternative is to make the $if$ condition at line 10 of Algorithm~\ref{Algorithm:GenAlg}
iterate over the mappings associated with each SNP.
While it would result in less I/O time, fixing the loop bound
according to the maximum number of mappings decreases performance
if the majority of SNPs have few associated mappings. This would also
complicate distinguishing between entries in our compact representation.

\section{Other Use Cases}
\label{sec:discuss}

Our OTP construction can also be adapted to other uses for one-time programs;
we provide the intuition below. We must also consider the monetary costs associated with adapting
programs into OTP boxes according to our design. If non-interactivity is not required, interactive
garbled circuit protocols may suffice.

\paragraph{Additional genomic tests.} Other tests are possible that operate on a
sequenced genome. Further, Bob may have multiple inputs to evaluate on a single
function. For instance, an individual may input two or more genomes for a
paternity test or a disease predisposition test that may also involve other
family members. This functionality can be easily added to the proposed scheme by
treating multiple sets of test data as single input, although it does not provide
privacy between family members (but provides privacy of the set from the vendor).

\paragraph{Temporary transfer of cryptographic ability.}
OTP lends itself naturally to the situation when one party
must delegate to another
the ability to encrypt/decrypt or sign/verify messages~\cite{goldwasserOTP}.
In this case, individual OTP boxes must be provisioned and given in advance to the designee,
with each box only capable of performing a single crypto-operation.
The cost could easily add up, but it might be acceptable for time-sensitive
or infrequent messages of high importance (such as military communications).
If messages are more frequent, then it may be worthwhile to consider a $k$-time
extension ($k > 1$) to OTP.
In either case, the designee is never given access to the raw private key.
Care must be taken to restrict the usable time of each box,
which can be realized by sealing an end date in addition to the one-timeness flag.

\paragraph{One-time proofs.}
As suggested by Goldwasser \etal~\cite{goldwasserOTP}, OTP allows witness-owners to go offline
after supplying a proof token to the prover. This proof token can
be presented to a verifier only once, after which it is invalidated.
We can certainly realize this functionality using our OTP boxes,
since proofs produced by our OTP are invalidated by nature of interactive
proof systems and may not be reused. Depending on the usage environment, using our OTP
box may or may not be cost-effective. While our implemented system may be
too costly to serve as subway tickets, the cost may be justified if our box
is used as an access-control mechanism to a restricted area.

\paragraph{Digital cash.}
As a one-time program, this was investigated
by Kitamura \etal~\cite{moneyOTP}, which used Shamir's secret sharing in place of OTMs.
We borrow their three-party scenario to reason about our own OTP system.
\begin{enumerate}
    \item The bank supplies OTP boxes with set dollar values.
    \item To make a payment, the user provides to the OTP box the shop's
hash of a newly generated random number.
    \begin{itemize}
        \item In TXT, the corresponding keys are selected.
        \item After reboot, the selected keys are input into the garbled circuit
program, which outputs a signature of the dollar-value concatenated with
the shop's hash value.
    \end{itemize}
    \item The shop verifies the signature.
    \item The shop requests cash from the bank using the signature.
\end{enumerate}
Unlike Kitamura \etal~\cite{moneyOTP}, we have a proper OTM in the form of the TPM.
A sealed flag value could enforce the one-timeness, preventing the user
from giving valid signatures for more than one shop input.
However, our scheme requires further modification to prevent double-spending,
as it is possible for two independent shop hash-values to be the same, in
which case the user can reuse the associated signature.
Furthermore, OTP for digital cash would not be feasible if the held dollar value is
less than the cost of the OTP box itself, unless the bank customer's
goal is untraceability.

\section{Table of SNPs on BRCA1 and their risk factors}
\label{sec:SNPTable}
\vspace{-10pt}

\begin{table}[]
	\footnotesize
	\centering
	\begin{tabular}{ | c | c | c | c | p{0.5cm}|}
		\hline
		SNP Reference Number & Position & Alleles & Risk Factor \\ \hline
		\multirow{4}{*}{rs41293463} & \multirow{3}{*}{43051071} & AT &  6 \\
		\cline{3-4}&& GG & 6   \\
		\cline{3-4}&& GT & 6 \\ \hline
		\multirow{3}{*}{ rs28897696 } & \multirow{2}{*}{43063903} & AA &  7 \\
		\cline{3-4}&& AC & 6   \\\hline
		\multirow{3}{*}{  rs55770810 } & \multirow{2}{*}{43063931} & CT &  5 \\
		\cline{3-4}&& TT & 5   \\ \hline
		\multirow{3}{*}{  rs1799966 } & \multirow{2}{*}{ 43071077 } & GG &  2  \\
		\cline{3-4}&& AG & 1.1   \\ \hline
		\multirow{4}{*}{rs41293455} & \multirow{3}{*}{ 43082434} & CG &  5 \\
		\cline{3-4}&& CT & 5   \\
		\cline{3-4}&& TT & 2 \\\hline
		\multirow{3}{*}{ rs1799950 } & \multirow{2}{*}{ 43094464 } & GG &  2  \\
		\cline{3-4}&& AG & 1.5   \\ \hline
		\multirow{1}{*}{rs4986850 } & \multirow{1}{*}{ 43093454 } & AA &  2  \\\hline
		\multirow{1}{*}{ rs2227945 } & \multirow{1}{*}{ 43092113 } & GG &  2  \\ \hline
		\multirow{2}{*}{ rs16942 } & \multirow{2}{*}{ 43091983} & AG &  2  \\
		\cline{3-4}&& GG & 2   \\\hline
		\multirow{1}{*}{  rs1800709} & \multirow{1}{*}{  43093010 } & TT &  2  \\ \hline
		\multirow{1}{*}{ rs4986852} & \multirow{1}{*}{ 43092412 } & AA &  2  \\ \hline
		\multirow{2}{*}{rs28897672} & \multirow{2}{*}{ 43106487} & GG &  4 \\
		\cline{3-4}&& GT & 4   \\ \hline
	\end{tabular}
	\vspace{5pt}
	\caption {SNPs on BRCA1 and their corresponding risk factors for breast cancer.}
	\label{Table:SNPsBRCA}
	\vspace{-15pt}
\end{table}

\noindent
The genetic instructions that determine development, growth and certain functions are carried on Deoxyribonucleic acid (DNA)~\cite{genetics}. DNA is in the form of double helix, which means that DNA consists of two polymer chains that complement each other. These chains consist of four nucleotides: Adenine (A), Guanine (G), Cytosine (C), and Thymine (T). Genetic variations are the reason that approximately $0.5\%$ of an individual's DNA is different from the reference genome.

SNP defines a position in the genome referring to a nucleotide that varies between individuals. Each person has approximately 4 million SNPs. Each SNP contains two alleles, which correspond to nucleotides

Table~\ref{Table:SNPsBRCA} lists the SNPs related to BRCA1 and the corresponding risk factors. The magnitude of risk factors ranges from 0 to 10~\cite{snpMagnitude}. A risk factor greater than 3 indicates a significant contribution of that particular allele combination to the overall risk of contracting breast cancer.

\section{Genomic Algorithm}
\label{sec:GeneticAlgo}
The genomic algorithm is shown in Algorithm~\ref{Algorithm:GenAlg}. Here, RF corresponds to the total risk factor for developing breast cancer. The "risk factors" file contains the associations between the observed SNP and the risk factor, while the "patient SNPs" file contains a patient's extracted SNPs.

\begin{algorithm}[ht] 
	\footnotesize
	\caption{Genomic Algorithm}
	\label{Algorithm:GenAlg}
	\begin{algorithmic}[1]
		\item[\textbf{Input:}] $Risk Factors, Patient~SNPs$
		\item[\textbf{Output:}] $RF$
		\Procedure{Genomic Algorithm}{$Risk Factors, Patient~SNPs$}
		\State $RF = 0$ 
		\For{\texttt{line in Patient~SNPs}}
		\State \texttt{SNP\_ID = SNP ID in line}
		\State \texttt{ALLELES = two alleles in line}
		\For{\texttt{line\_rf in Risk~Factors}}
		\State \texttt{SNP\_ID\_rf = SNP ID in line\_rf}
		\State \texttt{ALLELES\_rf = two alleles in line\_rf}
		\If{\texttt{SNP\_ID = SNP\_ID\_rf}}
		\If{\texttt{ALLELES = ALLELES\_rf}}
		\State RF = RF + risk factor in  line\_rf
		\Else
		\State RF = RF + 0
		\EndIf
		\Else
		\State RF = RF + 0
		\EndIf
		
		\EndFor
		\EndFor
		\State \textbf{return} $RF$
		\EndProcedure
	\end{algorithmic}
\end{algorithm}

\section{Porting effort}
\label{appendix:port}
To get started with an application based on our OTP (either variant), the very first step is always
creating the payload program, or if existent, porting it to a designated programming language.
In the case of the GC-based, rewriting in the Frigate-specific limited-C language is necessary.
For the TXT-only, the few technical tweaks such as PATA I/O with our added DMA support are seamlessly
transparent to the application developer, since they are only exposed like POSIX-like file operations,
similar to \emph{fopen{}}, \emph{fread{}} and \emph{fwrite{}}. We argue that regarding the status quo
of most existing OTP solutions, this process has to be (quasi-)manual in terms of programming language.
Therefore, we hope, as future work, to either have an automated framework for OTP-specific conversions
or (in the case of TXT-only) include a lightweight language environment.

\vspace{-5pt}
\section{SMM attacks and TPM relay attack} \label{sec:relay-smm-attacks}
\vspace{-5pt}
\bfhead{SMM attacks.} The System Management Mode (SMM) is a special
execution mode in modern x86 CPUs and considered having (informally)
the Ring minus 2 privilege, preempting virtually any other modes. Therefore,
although not recently, it was exploited~\cite{attack-txt} to interfere with TXT
execution. This attack assumes the compromise of the SMI (SMM Interrupt)
handler (which is difficult, but feasible in an ad-hoc manner), and during TXT
execution an SMI is triggered and the compromised handler comes in to
manipulate anything of the adversary's choice. However, in the case of our
OTP, we do not load any standard code that needs SMI and has it enabled
(like an OS or hypervisor); instead, our custom program for key selection or
OTP execution has SMI disabled from the first line of code (not to mention
containing any SMI trigger, e.g., writing to port 0xb2), and thus is not affected
by such attacks. Since TXT is exclusive, no other code can run in parallel.

Note that Alice no longer benefits from any attacks (e.g., stealing Bob's
input) due to loss of physical possession and network connectivity.
We exclude, for now, any potential threats from Intel ME (Management Engine)
which is referred to as Ring minus 3 and has a dedicated processor,
in the consideration that all rely on ad-hoc vulnerabilities and this topic
is still under open discussion~\cite{me}. 

\bfhead{TPM relay attack~\cite{cuckoo}.} A man-in-the-middle (MitM) attack
specifically targeting TPM-like devices impersonates and forwards requests to
a (remote) legitimate device, pretending its proximity or co-location on the same machine,
to either learn the secrets or forge authentication/attestation results. In the case of our OTP,
only Bob has physical possession and is motivated for such attacks. However,
since he cannot clone the TPM chip, whatever real traffic directed to the legitimate
one will cause irreversible effect (e.g., flipping the flag)
Note that his intension is not merely mimicking, which does not help. Also, we do not send
TPM commands in plaintext, except for ordinals and certain metadata.
Our ultimate argument is that, regardless of the lab effort we already exclude
in Section~\ref{sec:threat}, the integration of TPM in other microchips (e.g., SuperIO)
or an equivalent method will avoid exposing TPM pins for potential probing.

\end{document}